# A Novel Approach for Computing Dynamic Slices of Aspect-Oriented Programs

Abhishek Ray

School of Computer Engineering, Kiit University
Bhubaneswar, Odisha, India
ar_mmclub@yahoo.com

Siba Mishra

School of Computer Engineering, Kiit University
Bhubaneswar, Odisha, India
sibamishra@yahoo.co.in

Durga Prasad Mohapatra

Department of Computer Science & Engineering,
National Institute of Technology
Rourkela, Odisha, India
durga@nitrkl.ac.in

*Abstract*— We propose a dynamic slicing algorithm to compute the slices of aspect-oriented programs. We use a dependence based intermediate program representation called Aspect System Dependence Graph (AOSG) to represent aspect-oriented programs. Then, we propose the dynamic slicing algorithm for AOPs, which is an extended version of EMDS algorithm for object-oriented programs. Our algorithm is based on marking and unmarking of the edges of AOSG appropriately during run-time.

*Keywords-Aspect-oriented Programming; Aspect-oriented System Dependence Graph; Aspect; System Dependence Graph; Slice*

## I. INTRODUCTION

Program slicing [3, 4, 7, 14, 28, 29, 30, 31] is a practical disintegration methodology that omits program modules that are irrelevant to a particular process based on a criterion known as the slicing criterion [29, 30, 31]. The original program's semantics is projected through the computation of an executable program formed by the left over modules called a slice. Using this methodology, we can automatically determine the relevance of a module in a particular computation. The concept of static program slicing was introduced by Weiser [29, 30, 31]. A program $P$ can be sliced with respect to a "slicing criterion". A slicing criterion [29, 30, 31] consists of a pair $<s, v>$, where s is a program point and v is a subset of program variables. The parts of a program that have a direct or indirect effect on the values computed at a slicing criterion are called the program slice with respect to criterion. A static slice of a program P with respect to a slicing criterion $<s, v>$ is the set of all the statements of the program P that might affect the slicing criterion for any possible inputs to the program. Korel and Laski [14] introduced the concept of a dynamic program slicing. A dynamic slice contains only those statements that actually affect the slicing criterion for a given execution. Dynamic slicing [3, 14] is used to find the slice of a program with a certain input. This approach results in a possible smaller slice, because some options in the control flow might be eliminated. Program slicing has been found to be useful in a variety of applications such as debugging, program understanding, maintenance, testing and model checking and program comprehension.

Aspect-Oriented Programming (AOP) [1, 2, 5, 8, 13, 15, 17] is an emerging programming paradigm that was first published in June 1997 by Gregor Kiczale's team from Xerox Palo Alto Research Centre (PARC). The aim of AOP is an elegant, efficient and easily maintainable way to handle the parts of a program that are not easily addressable with traditional programming paradigms because these parts affects not only the single abstractable parts of the software but also affects the whole system, and making it difficult and inefficient to pack it into functional units. AOP is a new programming methodology which enables a clear separation of concerns in software applications. A concern in mean of software development can be understood as "a specific requirement that must be addressed in order to satisfy the overall system goal". Concerns can be subdivided in two subgroups:

- Core Concerns: They specify the main functionality of a software module, such as account withdrawing or button click behaviour.
- Cross-cutting Concerns: They specify the peripheral requirements that cross multiple modules, such as transaction handling, threading, authentication, logging, synchronization, security etc.

## II. ASPECTJ – AN AOP LANGUAGE

The most popular AOP language is AspectJ [1,2, 5,6,10,12,17]. AspectJ was created by Chris Maeda at Xerox Palo Alto Research Center (PARC). This is an aspect-oriented extension to Java programming language. In other words, we can say that AspectJ is compatible with Java platform. The recently released version of AspectJ (1.1) has the required maturity that the language and tools need to make it possible to work with large projects. It also features a compiler based on the industry-strength Java compiler that is a part of the Eclipse IDE and used by thousands of projects worldwide. An AspectJ program is divided into two parts: base code or non-aspect code and aspect code. The base code includes classes, interfaces and other standard Java constructs. The aspect code implements the crosscutting concerns in the program. Fig. 1 shows a sample AspectJ Program.

## III. SOME DEFINITIONS

In this section, we present some basic definitions, concepts, notations and terminologies associated with the intermediate program representation. These concepts and definitions are later used throughout the discussion in this paper. The basic concepts and definitions of def(Var), use(var), defSet(var) and useSet(var), (recentDef(var)), control dependence and data dependence are available in [22, 23]. However, here we present some of them for the sake of conciseness and completeness. In the rest of the paper, we use the terms *node* and *vertex* interchangeably.

**Definition 1 (Precise Dynamic Slice):** A dynamic slice is said to be precise if it includes only those statements that actually affect the value of a variable V at a statement S for the given execution.

**Definition 2 (Correct Slice):** A correct slice contains all the statements that affect the slicing criterion. A correct slice is imprecise if and only if it contains at least one statement that doesn't affect the slicing criterion.

**Definition 3 (def(Var)):** Let *Var* can be an instance variable in a class in an aspect-oriented program. A node *x* is said to be a def(Var) node if *x* represents a definition for the variable *Var*.

| Non-Aspect Code | Aspect Code |
|---|---|
| import java.util.*;<br>class prime<br>{<br>static int n;<br>1 public static void main(String args[])<br>{<br>2 n = Integer.ParseInt(args[0]);<br>3 if (isprime(n))<br>4 system.out.println("Is Prime");<br>else<br>5 system.out.println("Is not Prime");<br>}<br>6 static boolean isprime(int n)<br>{<br>7 for(int i=2;i<=n/2;i++)<br>{<br>8 if(n%i == 0)<br>9 return false;<br>}<br>10 return true;<br>}<br>} | l1 public aspect PrimeAspect<br>{<br>l2 public pointcut primeoperation(int n):<br>call(boolean prime isprime(int) &&<br>args(n));<br>l3 before (int n) : primeoperation(n)<br>{<br>l4 System.out.println("Testing the<br>prime no for :"+n);<br>}<br>l5 after (int n) returning(boolean result):<br>primeoperation(n)<br>{<br>l6 System.out.println("Showing the<br>prime status for :"+n);<br>}<br>} |

Figure 1. A Sample AspectJ Program.

**Definition 4 (defSet(var)):** The set defSet(var) denotes the set of all def(Var) nodes.

**Definition 5 (use(var)):** Let *var* be a variable defined in a class in an object-oriented program. A node *x* is said to be a use(Var) node, if and only if it uses the variable *var*.

**Definition 6 (useSet(var)):** The set useSet(var) denotes the set of all (use(var)) nodes.

**Definition 7 (RecentDef(var)):** For each variable var, recentDef(var) represents the node corresponding to the most recent definition of var with respect to some point s in an execution.

**Definition 8 (C-Node):** The C-Node maintains the logical connectivity among different weaving points. A C-Node doesn't represent any specific statement in the source code of a program. C-Nodes capture communication dependencies among the non aspect code and aspect code, Since C-Nodes are not mapped to any specific program statement, and we call them dummy nodes or logical nodes.

## IV. PROPOSED WORK

This section, discusses the construction of our proposed intermediate program representation Aspect System Dependence Graph (AOSG) and also the algorithm for computing the dynamic slice of AOP by using AOSG.

### A. Construction of AOSG

The construction of AOSG consists of the following steps:

*Construction of SDG for non-aspect code of the program:*

The system dependence graph (SDG) [11, 16] of an object-oriented program (OOP) or the non-aspect code of an aspect-oriented program (AOP) is a collection of method dependence graphs (MDGs). A method dependence graph (MDG) for a method contains vertices representing statements or predicates of the methods and edges representing data and control dependences among the statements. The construction of the complete SDG can be performed by connecting MDGs at call sites. Fig. 2 shows the SDG of non-aspect code of Fig. 1.

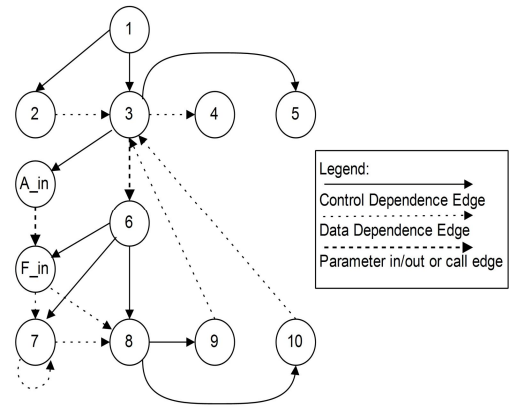

Figure 2. SDG for the non-aspect code given in Fig. 1.

*Construction of ADG for the aspect code of the program:*

The dependence graph for aspect code, called the Aspect Dependence Graph (ADG), is constructed by combining the advice dependence graph, introduction dependence graph, pointcut dependence graph and method dependence graph.

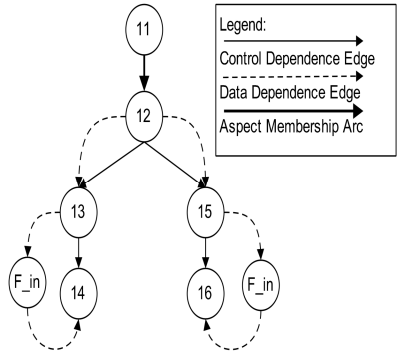

Figure 3. ADG for the aspect code given in Fig. 1.

The advice dependence graph and introduction dependence graph of advices, and introductions are similar to the MDG of a method. Pointcut dependence graph is used to represent a pointcut in an aspect. As Pointcuts declared in an aspect contain no body, so for each pointcut designator, the pointcut start vertex need to be constructed that will be the entry point to the pointcut. The ADG corresponds to the aspect code of Fig. 1 is shown in Fig. 3.

*Determination of aspect-membership arcs:*

The aspect-membership arcs are determined from the non-aspect code with the help of join-points. The join-points are well defined points along the execution of a program. By careful examination of join points declared in the pointcuts and their associated advice, the aspect-membership arcs can be determined to connect the SDG of the non-aspect code to the ADG of the aspect code. The aspect-membership arc is used to connect the SDG of base code with the ADG of aspect code.

*Connecting the SDG and the ADG:*

To construct the complete AOSG for an AspectJ program, the SDG of the non-aspect code and the ADG of the aspect code are connected by using aspect-membership arcs. C-Nodes are used for show the communication between the non-aspect and aspect codes. Actual and formal parameter vertices are connected by parameter arcs. The complete AOSG is shown in Fig. 4.

**B. Proposed Algorithm**

Before the execution of an aspect-oriented program (*P*), its Aspect System Dependence Graph is constructed statically once. During the execution of the program *P*, the algorithm marks and unmarks the executed edges of the AOSG appropriately. The pseudo code of our proposed algorithm is as follows:

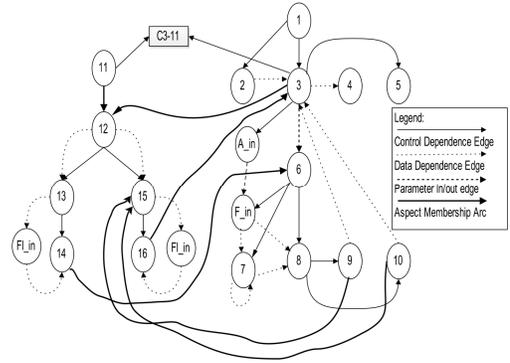

Figure 4. AOSG for AspectJ program given in Fig. 1.

**Stage 1: Constructing Static Graph**
1)  AOSG Construction
    a.  Add control dependence edge
        for each test (predicate) node *u*, do
            for each node x in the scope of *u* do
                Add control dependence edge (*u*, *x*) and mark it.

    b.  Add data dependency edges
        for each node *x*, do
            for each variable var used at x, do
                for each reaching definition u of var do
                    Add data dependence edge (*u*, *x*) and unmark it.

    c.  Add Aspect Membership edges
        for each join points *j*, do
            Add an Aspect Membership edge(*j*, *x*) for every node *x* that is call dependent on *j* and unmark it.

**Stage 2: Managing the AOSG at run time**
1)  Initialization: Do the following before execution of the program P.
    (a) Unmark all the edges.
    (b) Set dslice(*u*) = ∅ for every node *u*
    (c) Set RecentDef(*var*) = NULL for every variable *var* of the program *P*.
2)  Perform steps (3) and (4) until one condition in step (4) is satisfied.
3)  Runtime Updations: During runtime, carryout the following after each statement *S* of the program *P* is executed. Let node *u* in AOSG correspond to the statement *S* in Program *P*.
    (a) For every variable *var* used at node *u*, update
        dslice(u, var)= {$x_1$, $x_2$, ..., $x_k$} ∪ dslice($x_1$, var)
            ∪ dslice($x_2$, var) ∪...∪ dslice($x_k$, var)
    (b) For every variable *var* used at node *u* do the followings:
        i.    Unmark the marked dependence edges, if any associated with the variable *var*, which may have

been marked by the previous execution of the node *u*.

   ii.   Mark the dependence edge $(x,u)$ where $x =$ RecentDef(*var*).

(c) If *u* is a method entry vertex, then do the following

   i.   Unmark all the marked edges including call edges, parameter edges and summary edge, corresponding to the previous execution of the node *u*.

   ii.   Mark the corresponding parameter edges between the actual parameter vertices and formal parameter vertices.

(d) If *u* is a vertex representing the operator *new* then do the following

   i.   Mark actual_in and actual_out edges associated with *u* corresponding to the present execution of *u*.

   ii.   Mark the method entry edge of the corresponding construction method for the present execution of *u*.

   iii.   Mark formal_in and formal_out edges associated with the method entry edge of the constructor method.

(e) If *u* is a pointcut start vertex, then do the following

   i.   Mark actual_in and actual_out edges associated with *u* corresponding to the present execution of *u*.

   ii.   Mark the corresponding advice edge for the present execution of *u*.

   iii.   Mark formal_in and formal_out edges associated with the advice edge of the pointcut.

4) Slice look up

(a) If a slicing command *<u, var>* is given carry out the following:

   i.   Look up dslice(*u,var*) for variable var for the content of the slice.

   ii.   Display the result and exit.

(b) If the program is not terminated, go to step 2.

The space complexity of our proposed algorithm is $O(n^2)$, where, *n* is the number of statements of the program. The total execution time requirement for computing and updating information corresponding to an execution of a statement is $O(n^2)$. Hence, the run-time complexity of our algorithm for computing the dynamic slice, for the entire execution of the program is $O(n^2)$.

*C. Working of Algorithm*

Consider the sample AspectJ program given in Fig. 1. The SDG corresponding to the non-aspect code or base code is given in Fig. 2. The ADG corresponding to the aspect code is shown in Fig. 3. The complete AOSG is shown in Fig. 4.

Suppose we want to compute dynamic slice for the slicing criterion <16, n> of Figure 1. The dynamic slice is computed by dslice(16,n) = {15} $\cup$ dslice(15, n) . By evaluating the expression recursively the final dynamic slices are obtained. The group of recursive equations are given below.

dslice(15,n)  = {9,10,12} $\cup$ dslice(9,n)$\cup$ dslice(10,n)$\cup$ dslice(12,n)

dslice(9,n)   = {8} $\cup$ dslice(8,n).

dslice(10,n)  = {8} $\cup$ dslice(8,n).

dslice(12,n)  = {3,11} $\cup$ dslice(3,n) $\cup$ dslice(11,n).

dslice(8,n)   = {6,7} $\cup$ dslice(6,n) $\cup$ dslice(7,n).

dslice(3,n)   = {9,10,2,1,16} $\cup$ dslice(9,n) $\cup$ dslice(10,n) $\cup$ dslice(2,n) $\cup$ dslice(1,n) $\cup$ dslice(16,n).

dslice(11,n)  = {Ø} $\cup$ dslice{Ø}.

dslice(6,n)   = {3,14} $\cup$ dslice(3,n) $\cup$ dslice(14,n).

dslice(7,n)   = {6} $\cup$ dslice(6,n).

dslice(2,n)   = {1} $\cup$ dslice(1,n).

dslice(1,n)   = {Ø} $\cup$ dslice{Ø}.

dslice(16,n)  = {15} $\cup$ dslice{15}.

dslice(14,n)  = {13} $\cup$ dslice{13}.

dslice(13,n)  = {12} $\cup$ dslice{12}.

Thus, by using the definition of dslice(16, n) and considering group of recursive equations, the data structure dslice(16,n) set gets updated to {16,15} $\cup$ {15,9,10,12} $\cup$ {9,8} $\cup$ {10,8} $\cup$ {12,3,11} $\cup$ {8,6,7} $\cup$ {3,9,10,2,1,16} $\cup$ {11, Ø} $\cup$ {6,3,14} $\cup$ {7,6} $\cup$ {2,1} $\cup$ {1, Ø} $\cup$ {16,15} $\cup$ {14,13} $\cup$ {13,12} which equals to {16, 15, 9, 10, 12, 8, 3, 11, 6, 7, 2, 1, 14, 13}.

The statements included in the dynamic slices are shown in rectangular boxes in Fig. 5.

Figure 5: The dynamic slices of the program given in Fig. 1 for the slicing criterion <16, n>

V. IMPLEMENTATION

In this section, we briefly describe the implementation of our algorithm. The main motivation for our implementation is to verify the correctness and the preciseness of our algorithm. We have named our slicing tool as Dynamic Edge Marking Slicing Tool (DEMST) for aspect-oriented programs. In the next section we present an overview and working of the slicing tool.

The working of the slicing tool is schematically shown in Fig. 6. The arrows in Fig. 6 show the data flow among the different blocks of the tool. The blocks shown in rectangular boxes represent executable components and the blocks shown in ellipses represent passive components of the slicing tool.

A program written in AspectJ is given as input to DEMST. The overall control for the slicer is done through with the help of a *Graphical User Interface* (GUI). The *Slicer* takes user input from the GUI, interacts with other relevant components to extract the desired results and returns the output back to the GUI. The *ajc Preprocessor* reads the aspect headers preprocesses the Java source code and determines where the given aspects must take effect. It then weaves the aspects into the code, generating new source files. After completion of weaving process, the control is passed to the Lexical Analyzer. The *lexical analyzer* component reads the new source files of the input AspectJ program from the *ajc preprocessor* and breaks it into tokens for the grammar expressed in the parser. When the lexical analyzer component encounters a useful token in the program, it returns the token to the parser describing the type of encountered token. The *parser and semantic analyzer* component, functions as a state machine. The parser takes the token given by the lexical analyzer and examines it using the grammatical rules laid for the input programs. *The semantic analyzer component* captures the following important information of the program.

- For each node *w* of the program

1) the lexical successor and predecessor vertices of *w*,
2) the sets of variables defined and used at vertex *w*,
3) the type of the vertex: assignment or test or method call or return etc.

The input program code is appropriately instrumented with the help of code instrumentor and semantic analyzer component, so as to facilitate computation of dynamic slices and to update other associated run-time data structures after execution of each statement. The lexical analyzer component, parser and semantic analyzer component provides the necessary information to the *AOSG Constructor* component. Also, the *Program Analysis Block* provides necessary information and data structures to the *AOSG Constructor Block*. The *AOSG Constructor* component first constructs the CFG and the post-dominator tree of the program using the basic information provided by the lexical and semantic analyzer components. The inter-statement control dependences are captured using the CFG and the post-dominator tree. Then, it constructs the AOSG of input program along with all the required information to compute slices and stores it in appropriate data structures.

The overall control for the *Slicer* is done through the help of a GUI. The *Slicer* takes user input from the GUI, interacts with other relevant components to extract the desired results and returns the output back to the GUI. The slicer component can use or update the AOSG and its associated data structures, as and when required. The graphical user interface (GUI) functions as a front end to our slicing tool. The GUI block encapsulates the complexities involved in the functioning of all the other blocks by supporting user-friendly interface to the user.

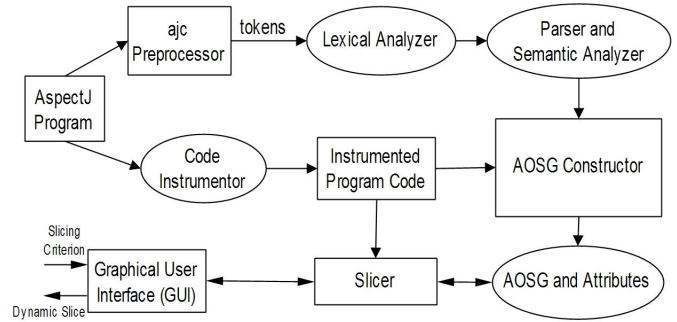

Figure 6: Schematic representation of working of the slicer

## VI. COMPARISION WITH RELATED WORK

Many works have been reported on slicing of procedural and object-oriented programs [9, 11, 19, 20, 22, 23, 24, 32, 33]. But, few works have been reported on slicing of AOPs [21, 25, 27, 34, 35]. Slicing aspect-oriented programs presents some challenges which are not encountered in traditional program slicing. The existing slicing algorithms for procedural and object-oriented programs cannot be applied directly to AOP, due to the presence of special features in aspect-oriented programs like aspects, pointcuts, join-points, advices etc. which needs a lot of investigation and new ideas in order to achieve a high-level of accuracy in computing dynamic slices.

Zhao [34] was first to develop the Aspect-oriented System Dependence Graph (ASDG) to represent aspect-oriented programs. The ASDG is constructed by combining the SDG [16] of non-aspect code, the aspect dependence graph (ADG) of aspect code and some additional dependence arcs used to connect the SDG and ADG. Then, Zhao used the two-phase slicing algorithm [16] proposed by Larsen and Harrold [16] to compute static slice of aspect-oriented programs. The main contribution is new dependence-based representation on which static slices of aspect-oriented program can be computed efficiently. The ASDG proposed by Zhao [34] for the example program of Fig. 1 is shown in Fig. 7. The limitations of this approach is that the concepts of pointcuts has not addressed clearly, the size of the ASDG is critical for applying to the practical software development and the proposed algorithm is not concrete in nature.

Later, SDG construction algorithm proposed by Zhao and Rinard [35] provides an efficient way for addressing the concepts of pointcuts. The SDG construction algorithm [35] is more concrete in nature, and can represent many aspect-oriented constructs. This algorithm produces precise static slices. Fig. 8 shows the SDG of the program given in Fig. 1 as proposed by Zhao and Rinard [35]. But the disadvantage of this SDG is that the weaving process is not represented correctly. The use of this SDG to compute dynamic slices results in missing of some statements.

Braak [27] extended the ASDG proposed by Zhao [34, 35] to include inter-type declarations or introductions in the graph. Each inter-type declaration was represented in the form of a field or a method as a successor of the particular class. Then, Braak [27] used the two-phase slicing algorithm of Horwitz et al. [11], to find the static slice of an aspect-oriented program. Though this technique is very efficient, but it can't be used for computing dynamic slices.

All the above mentioned techniques [27, 34, 35] provides a mean for the computation of static slices of an AOP. Program slicing techniques have been found useful in applications such as program understanding, debugging, testing, software maintenance and reverse engineering etc. Particularly, dynamic program slicing used in interactive applications such as debugging and testing of programs. Therefore, the dynamic slicing techniques need to be efficient. This requires developing efficient dynamic slicing algorithms as well as suitable intermediate representations for representing aspect-oriented programs.

With the above goals and motivations, Mohapatra et al. [21] proposed a new intermediate representation and named the program representation as Dynamic Aspect-Oriented Dependence Graph (DADG).

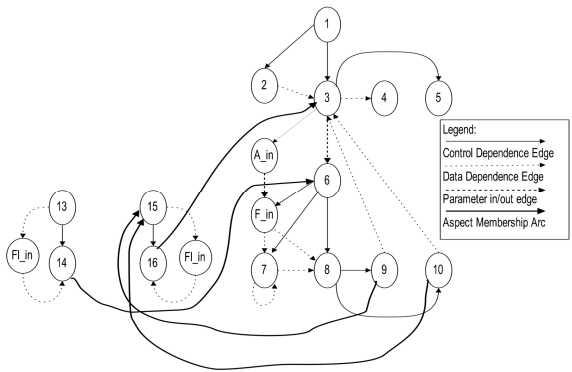

Figure 7: ASDG of the program given in Fig. 1, as proposed by Zhao [34]

They also proposed a dynamic slicing algorithm for aspect-oriented programs [21]. A trace file is used to store the execution history of the source code. The limitation of this approach [21] is that, it stores each occurrence of a statement in an execution trace which incurs lots of spaces and time yielding more time and space complexities.

Mohapatra et al. [25] also proposed a node marking dynamic slicing (NMDS) algorithm for computing the dynamic slice of an AOP. This algorithm is a quite efficient dynamic slicing technique, compared to the TBDS algorithm [21]. This is because, in this algorithm Extended Aspect-oriented System Dependence Graph (EASDG) is used to represent the aspect oriented program, and no trace file is used for storing the execution history.

Our approach is different from all other existing methods [21, 25, 27, 34, 35]. First, we construct the Aspect System

Dependence Graph (AOSG) to represent aspect-oriented programs. We have used C-Node to connect the non-aspect and aspect code

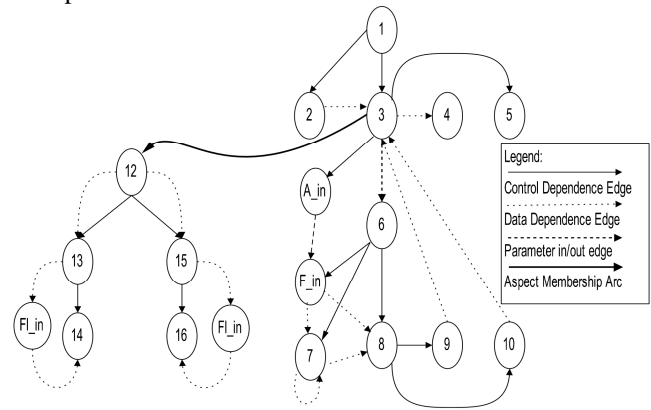

Figure 8: SDG of the program given in Fig. 1, as proposed by Zhao and Rinard [35].

Our proposed intermediate representation (AOSG) differs from the existing approaches in the following ways:

- The pointcuts in the AspectJ program are represented correctly and clearly.
- The mechanism of connecting non-aspect code and aspect code is explained clearly, during the construction of AOSG with the help of C-Nodes.
- The size of AOSG is not critical, thus our proposed AOSG can be applied to practical software development environment.
- Our proposed AOSG can represent many more aspect-oriented constructs such as aspects, pointcuts, joinpoints, advice precedence, introductions, scope and aspect-inheritance.
- There is no need for creating any new nodes in AOSG during run time.

After constructing the AOSG, our proposed algorithm marks and unmarks the edges of AOSG when dependencies arises and ceases during the run time, to compute the dynamic slice. Our algorithm does not require any trace files to store the execution history.

## VII. CONCLUSION

In this paper, we have proposed an algorithm for computing dynamic slices of aspect-oriented programs. Our algorithm used Aspect System Dependence Graph (AOSG) as the intermediate representation. Our proposed algorithm is based on marking and unmarking of the edges of AOSG as and when the dependencies arise and cease during run-time. The advantage of our algorithm is that, it does not create any additional nodes during run-time. Another advantage of our algorithm is that a slice is available even before a request for a slice is made. This reduces the response time of slicing commands.

## AUTHORS PROFILE

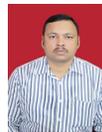

**Abhishek Ray** received his M. E. from Regional Engineering College (now NIT), Rourkela. He served as a faculty of the Department of Computer Science and Engineering at Gandhi Institute of Engineering & Technoly, Gunupur, Odisha, from 1998 to 2005. He joined as a faculty in School of Computer Engineering at KIIT University, Bhubaneswar, Odisha in 2005, where he is now Associate Professor. His research interests include software engineering, real-time systems, automata theory and compiler design. He has published five papers in these fields. Mr. Ray has been teaching automata theory and compiler design to UG and PG students at KIIT University, Bhubaneswar for the six years.

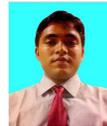

**Siba Mishra** received his M.Tech from KIIT University, Bhubaneswar, Odisha, India. His research interests include software engineering, automata theory, discrete mathematics.

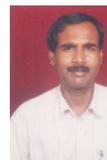

**Durga Prasad Mohapatra** received his Ph. D. from Indian Institute of Technology Kharagpur and M. E. from Regional Engineering College (now NIT), Rourkela. He joined the faculty of the Department of Computer Science and Engineering at the National Institute of Technology, Rourkela in 1996, where he is now Associate Professor. His research interests include software engineering, real-time systems, discrete mathematics and distributed computing. He has published more than thirty papers in these fields. Dr. Mohapatra has been teaching software engineering and discrete mathematics to UG and PG students at NIT Rourkela for the past ten years. He has received many awards including Young Scientist Award for the year 2006 by Orissa Bigyan Academy, Prof. K. Arumugam award for innovative research for the year 2009 and Maharasthra State National Award for outstanding research for the year 2010 by ISTE, NewDelhi. He has also received three research projects from DST and UGC, Government of India. Currently, he is a member of IEEE. Dr. Mohapatra has co-authored the book *Elements of Discrete Mathematics: A computer Oriented Approach* published by Tata Mc-Graw Hill.